\documentclass[%
 reprint,
 superscriptaddress,
 amsmath,amssymb,
 aps,
 prmaterials,
]{revtex4-2}

\usepackage{graphicx}
\usepackage{dcolumn}
\usepackage{bm}

\usepackage[utf8]{inputenc}
\usepackage[T1]{fontenc}
\usepackage{mathptmx}
\usepackage{etoolbox}
\usepackage{float}
\usepackage{hyperref}

\begin{document}

\title{Nonlinear Photocurrent Spectroscopy and Polarization-Tunable Shift Current in the Layered Semiconductor $\mathrm{CuScP_2S_6}$}
\author{Alexander M. Blackston}
    \affiliation{ 
Dept. of Material Science and Engineering, The Ohio State University, Columbus, Ohio, 43210, USA
}%
\author{Md Kazi Rokunuzzaman}%
    \affiliation{ 
Dept. of Material Science and Engineering, The Ohio State University, Columbus, Ohio, 43210, USA
}%

\author{Ryan P. Siebenaller}
    \affiliation{ 
Dept. of Material Science and Engineering, The Ohio State University, Columbus, Ohio, 43210, USA
}%
    \affiliation{ 
Foundational Technologies Directorate, Air Force Research Laboratory, Wright-Patterson Air Force Base, OH, 45433, USA
}%

\author{Shashu Tomar}%
    \affiliation{ 
Dept. of Physics, The Ohio State University, Columbus, Ohio, 43210, USA
}%

\author{Enam Chowdhury}
    \affiliation{ 
Dept. of Material Science and Engineering, The Ohio State University, Columbus, Ohio, 43210, USA
}%
    \affiliation{ 
Dept. of Physics, The Ohio State University, Columbus, Ohio, 43210, USA
}%
\author{Michael A. Susner}
    \affiliation{ 
Foundational Technologies Directorate, Air Force Research Laboratory, Wright-Patterson Air Force Base, OH, 45433, USA
}%
\author{Roberto C. Myers}
\email{myers.1079@osu.edu} 
    \affiliation{ 
Dept. of Material Science and Engineering, The Ohio State University, Columbus, Ohio, 43210, USA
}%
    \affiliation{ 
Dept. of Physics, The Ohio State University, Columbus, Ohio, 43210, USA
}%
\affiliation{Department of Electrical and Computer Engineering, The Ohio State University, Columbus OH 43210, USA}

\date{\today}

\begin{abstract}
The class of layered metal thiophosphates and selenophosphates are emerging as promising candidates for non-linear photonic and optoelectronic applications. Here we report one-photon absorption (1PA) and two-photon absorption (2PA) induced photocurrents in $CuScP_2S_6$, establishing it as an optoelectronically-active non-centrosymmetric semiconductor. Spectrally-resolved 1PA and 2PA photoresponsivity paired with photoluminescence and differential reflectance measurements reveal a band gap of 2.35 eV, several higher energy interband transitions, and sub-band gap multiphoton excitonic transitions. Vertical graphite/$CuScP_2S_6$/graphite devices exhibit zero-bias photocurrent with a polarity controlled by both photon energy and optical polarization orientation within the ab-plane. This sign-dependence combined with a characteristic power-law transition from linear to square-root scaling are consistent with a shift-current origin to this bulk photovoltaic effect (BPVE). These findings establish $CuScP_2S_6$ as a multi-functional platform for polarization-sensitive non-linear optoelectronics.
\end{abstract}
\maketitle
\section{\label{sec:intro} Introduction}

Recent studies of $CuScP_2S_6$ report its electrochemical energy storage capabilities \cite{oliveira_2d_2023} and high-\textit{k} dielectric behavior, including light-induced random telegraph noise in gated heterostructures \cite{ghosh_light-induced_2026}. Here we explore its intrinsic optoelectronic transport and photoconductive mechanisms using one-photon absorption (1PA) and two-photon absorption (2PA) photocurrent spectroscopy. Using lateral devices under bias, we resolve the 1PA photoresponsivity (PR) spectrum alongside a 2PA PR spectrum that exhibits narrow-linewidth PR peaks attributed to multiphoton excitonic transitions. Strong 2PA PR in $CuScP_2S_6$ can be explained by a recently reported inversion symmetry breaking (space group Cc) as demonstrated via strong second harmonic generation (SHG) and X-ray diffraction (XRD) Rietveld refinement methods.\cite{siebenaller_second_2024, siebenaller_nonlinear_2025} The 1PA and 2PA PR spectra analyzed alongside spectrally resolved photoluminescence and differential reflectance reveal $CuScP_2S_6$ to possess a direct band gap ($E_g$) of $\approx2.35 $ eV and higher energy optically active electronic transitions that allow for photocurrent generation into the deep UV, solar-blind spectral region out to at least 5.0 eV (248 nm). Additionally, two highly emissive deep level defect states are measured at $\approx1.3$  eV and $\approx2.0$ eV. These results confirm $CuScP_2S_6$ as a wide-bandgap semiconductor with rich nonlinear functionality.

Additionally, $CuScP_2S_6$ exhibits a zero-bias photocurrent enabled by the bulk photovoltaic effect (BPVE) in vertical heterostructures. BPVE is a bulk material process where photocurrents originate from a shift current, a phenomenon rooted in the quantum geometry of the material.\cite{ tanShiftCurrentBulk2016, pusch_energy_2023, aftab_engineering_2025, dai_recent_2023} Shift currents are expected to arise nearly instantaneously, occurring at the time scale of the underlying quantum transitions, opening up the potential for applications in ultrafast sensing.\cite{sotome_spectral_2019,he_ultrafast_2024} Since BPVE is a second-order non-linear effect, it only arises in materials with broken inversion symmetry and is tied to the lattice geometry and its associated Berry connection vector field. 

The direction of the BPVE response is contained within the shift vector, \textit{R}, which "describes the change in position that occurs as an electron absorbs a photon."\cite{sipeSecondorderOpticalResponse2000} The direction of this displacement depends on the transition dipole phase derivative, $\nabla_k \varphi^{a}_{vc}$, which is polarization axis ($a$) dependent. A second term controls the magnitude of the electron displacement, the change in Berry connection between the final conduction band ($c$) and initial valence band ($v$) states, $\Delta\chi_{cv}= \chi_{c}(k)-\chi_{v}(k) $, which is energy dependent, such that $R=\nabla_k \varphi^{a}_{vc}+\Delta\chi_{cv}$. \cite{tanShiftCurrentBulk2016}

Our spectral and polarization resolved BPVE photocurrent measurements show the shift current contribution to be dynamically switchable via photon energy and optical polarization, which is consistent with a polarization tunable $\nabla_k \varphi^{a}_{vc}$ and an energy dependent $\Delta\chi_{cv}$. These measurements suggest $CuScP_2S_6$ to be a promising platform for future ultrafast, polarization-sensitive applications. This work transitions $CuScP_2S_6$ from a passive dielectric or electrochemical material into an active platform for quantum-geometric optoelectronics.

\section{\label{sec:exp} Experimental Details}

\subsection{\label{sec:synth}Sample Synthesis}

To synthesize $CuScP_2S_6$ crystals, elemental Cu and Sc (Alfa Aesar, Puratronic) are mixed together with a pre-reacted $P_2S_5$ mixture (reacted from the elements at $330^\circ$ C for 10 hours) in the ratio Cu:Sc:$P_2S_5$ 1:1:3 in a 5 mL Canfield crucible set.\cite{noauthor_canfield_nodate} The crucible set is inserted into a quartz ampoule, sealed under 1/3 atm Ar, placed into a box furnace, heated to $650^\circ$ C at $20^\circ$ C/hour, held at that temperature for 12 hours, and slow cooled to $350^\circ$ C at a rate of $60^\circ$ C/hour, following previous procedures. \cite{chica_p2s5_2021} Once this final temperature is reached, the ampoule is placed upside-down in a centrifuge and spun  to high speed to decant the flux through the alumina frit. After cooling to room temperature, the assembly is opened to isolate the flakes. Chemical composition is verified with a Hitachi TM 4000 Plus mated with an Oxford Aztec One Xplore30 Compact EDS system; all compositions are found to be stoichiometric, within error. The structure is probed via X-ray diffraction using a Malvern PanAlytical Empyrian X-ray Diffractometer with Cu k-$\alpha$ radiation together with a two-bounce Ge (220) monochromator.\cite{siebenaller_nonlinear_2025}

\subsection{\label{sec:optics} Raman, Photoluminescence, and Reflection Spectroscopy}

Raman Spectroscopy measurements are performed on a Renishaw inVia Confocal Raman Microscope at room temperature using a 514 nm CW laser kept at 0.03 mW to excite the phonon modes. The laser is focused on the sample using a long working distance 50x objective with 0.5 N.A., and the spectrum is collected for 30 s with four accumulations in a reflection geometry using an 1800 l/mm grating.

Photoluminescence spectroscopy measurements are performed on an exfoliated $CuScP_2S_6$ flake in a cryostat at 150 K. A wavelength tunable pulsed femptosecond laser "Coherent Chameleon Ultra II" set to 800 nm is used to drive a third harmonic generation (THG) crystal at 266 nm using a commercially available harmonic box setup "Chameleon VUE harmonics." THG from the setup is focused onto the sample using a 40x UV enhanced aluminum reflective objective. A linear polarizer placed between the oscillator and harmonic box is rotated to attenuate the total THG power at the objective to 3 mW. PL is collected in a reflective geometry and the THG used to pump the PL is filtered out using a combination of a 266 nm long pass dichroic filter and a 280 nm long pass colored glass filter. The PL spectrum is resolved using a Princeton Instruments Acton SP2500 spectrometer paired with a Princeton Instruments Pixis 400, 1340 x 400 pixel array CCD camera.

Differential reflectance measurements are carried out using an ultra-stable deuterium-tungsten halogen light source (Ocean Optics DH-2000) paired with a 40x UV enhanced aluminum reflective objective. A 50 $\mu m$ core diameter optical fiber is placed in the image plane to sample the reflectance spectrum from an ~2 $\mu m$ diameter spot.\cite{frisenda_micro-reflectance_2017}

\subsection{\label{sec:fab}Device Fabrication}

Photocurrent devices are fabricated using a PDMS based dry transfer technique to transfer $CuScP_2S_6$ and graphite flakes as required. Maskless alignment photolithography is used to fabricate a mask using AZ NLOF 2020 photoresist through which Ti/Au contacts are deposited using electron beam evaporation. Metal lift-off is then completed using an N-Methyl-2-pyrrolidone (NMP) based photoresist solvent. 

\subsection{\label{sec:PR}1PA and BPVE Photoresponsivity (PR) Spectroscopy}

Linear single photon absorption (1PA) photocurrent measurements are carried out in a lateral geometry $CuScP_2S_6$ device using a monochromated laser driven xenon plasma broadband CW light source "MountainSource-Hyperchromator" which is focused to excite photocurrent in the depletion region near a positively biased electrode. A slit at the export port of the monochromator is narrowed to ensure that the full-width-half-max (FWHM) associated with the linewidth of the monochromated light is <1.75 nm across the entire spectrum used for the measurement. ( Fig. \ref{fig:supfig4}) The light source is focused onto the sample using a 40x UV enhanced aluminum reflective objective to mitigate chromatic aberration effects. The device is biased at 3V and photocurrent is measured in 2 nm steps using a Keithley 2636b Sourcemeter. The photocurent spectrum is normalized by the power spectrum that is obtained separately under the same condition after the objective using a calibrated power meter (Thorlabs S130VC), giving the PR spectrum. 

The BPVE PR spectrum is obtained from an unbiased vertical graphite-$CuScP_2S_6$-graphite device using the same setup and procedure as for the 1PA PR measurement. A linearly polarized 405 nm laser diode is further linearly polarized using a calcite polarizer and then rotated using a $\lambda/2$ wave-plate to obtain polarization resolved PR. A second linear polarizer is placed after the $\lambda/2$ wave-plate to obtain power dependent photocurrent measurements along the $\theta=0^\circ$ and $\theta=90^\circ$ polarization orientations. The polarization extinction ratio (PER) is measured to be 9000:1 ( Fig. \ref{fig:supfig6})

\subsection{\label{sec:2PR}2PA Photoresponsivity (PR) Spectroscopy}

A wavelength tunable pulsed femtosecond laser (Coherent Chameleon Ultra II) is used to excite 2PA photocurrent across a 3 V biased lateral geometry $CuScP_2S_6$ device from 680 nm to 1080 nm in 2 nm steps. Similar to the 1PA and BPVE measurement, the laser is focused using a 40x UV enhanced aluminum reflective objective and the photocurrent is measured with a Keithley 2636b Sourcemeter. The photocurent spectrum is normalized by the power spectrum that is obtained separately after the objective using the calibrated power meter, giving the PR spectrum. ( Fig. \ref{fig:supfig5})

\section{\label{sec:results} Results and Discussion}

\subsection{\label{sec:1PR} Linear Optical and Photocurrent Spectroscopy}

\begin{figure*}[t]
    \includegraphics[width=6.75in]{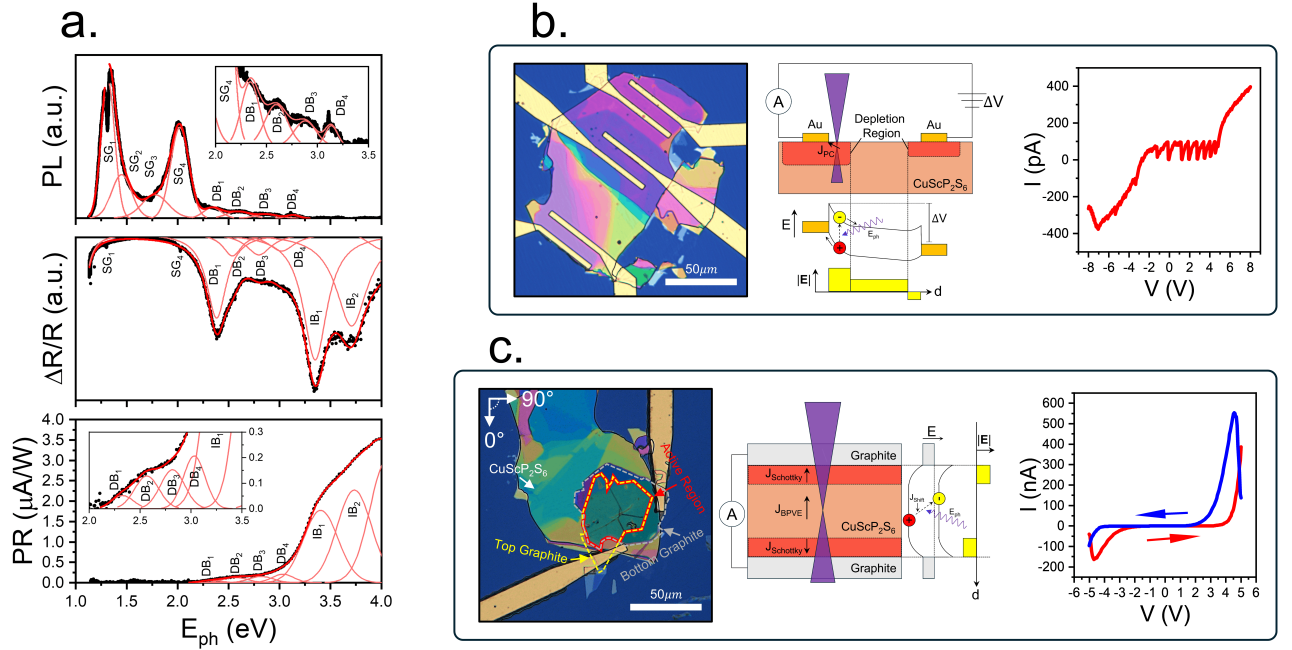}
    \caption{\textbf{a.)} \textit{Top}- PL spectrum collected at 150 K where fitted emission peaks have been labeled, \textit{Middle}- $\Delta R/R$ spectrum collected from lateral device active region where fitted reflectance peaks have been labeled, \textit{Bottom}- 1PA PR spectrum where PR peaks have been fitted and labeled. \textbf{b.)} \textit{Left}- microscope image of lateral photocurrent device, \textit{Center}- lateral device schematics with band diagram overlay,  \textit{Right}- IV characteristics measured across lateral device. \textbf{c.)} \textit{Left}- microscope image of vertical device with active region outlined, \textit{Center}- vertical device schematics with band diagram overlay, \textit{Right}- IV characteristics measured across vertical device.}
    \label{fig:fig1}
\end{figure*}

Mechanical exfoliation is employed to prepare thin chemical vapor transport (CVT) grown $CuScP_2S_6$ flakes and optical characterization is carried out via room temperature Raman ( Fig. \ref{fig:supfig1}) and low temperature (150 K) photoluminescence (PL) spectroscopy. (Figure \ref{fig:fig1}a- top)  Raman and PL spectra show features that replicate previous studies\cite{ghosh_light-induced_2026,oliveira_2d_2023}, confirming the flakes to be phase pure and suitable for device fabrication. Additionally, a Gaussian fit routine is employed to fit the PL spectrum to approximate the underlying electronic transition energies. (Table 1) Lateral geometry photocurrent devices are fabricated via electron beam evaporation of Ti/Au (5 nm/ 800 nm) contact electrodes spaced 10 $\mu m$ apart. (Figure \ref{fig:fig1}b-left) Atomic force microscopy is used to measure the thickness of the $CuScP_2S_6$ flake to be $\approx$ 200 nm.  Current-voltage (IV) measurements obtained across the device by sweeping voltage from -8 V to 8 V show behavior consistent with Schottky contact formation and strong defect trapping as evidenced by significant divergence from the exponential I-V relationship associated with ideal diode behavior and by the initial brief ramp up in current observed between -8 V and -7 V. (Figure \ref{fig:fig1}b- right) This non-instantaneous rise to full injection current levels and the observed current oscillation obtained between -4 V and 4 V are consistent with long time constant deep level trap state charging and discharging. This mirrors recent reports where trap states in $CuScP_2S_6$ are proposed to explain the large light-induced random telegraph noise in $CuScP_2S_6$ gated hetero-structures \cite{ghosh_light-induced_2026} as well as the known inclination for related metal thiophosphates, e.g. $CuInP_2S_6$ and $AgScP_2S_6$, to form sulfur vacancies. \cite{ mushtaqAgScP2S6VanWaals2023a, zhang_colossal_2023, yu_few-layered_2021}  Differential reflectance ($\Delta R/R$) measurements are obtained from the active device region between the two electrodes at zero bias to provide a baseline, all-optical measurement of the electronic structure of $CuScP_2S_6$ on the same flake used for photocurrent measurements. (Figure \ref{fig:fig1}a- middle) A best fit for the $\Delta R/R$ spectrum is obtained by applying a Lorentzian-based fitting routine which reveals reflectance peaks at energies that correlate well with those observed in the PL spectrum.

To collect photocurrent from the lateral device, a 3 V bias is applied between the electrodes and monochromated light from a laser driven xenon plasma lamp is focused to a diffraction limited spot near the electrode that is positively biased. (Figure \ref{fig:fig1}b-center) An increased photocurrent is obtained for this configuration compared to focusing the light near a negatively biased electrode, an indication of n-type Schottky contact formation. Using this excitation geometry, the linear, i.e. single photon absorption driven, photocurrent spectrum is collected and then normalized by the power spectrum of the light source to give the photoresponsivity (PR) spectrum.  A Gaussian fit routine, parametrized based on the electronic transitions observed in the PL and $\Delta R/R$ spectra, is applied and achieves a close fit. The fitted transition peaks show close agreement with those fitted in the PL and $\Delta R/R$ spectra.

\begin{table*}
\begin{ruledtabular}
\begin{tabular}{clccc}
Label & Description & PL Peak Energy & $\Delta R/R$ Peak Energy & PR Peak Energy\\
\colrule
\colrule

$SG_1$ & Defect trap state mediated transition & 1.32 & 1.30 & ---\\
\colrule
$SG_2$ & Defect phonon side band & 1.45 & --- & ---\\
\colrule
$SG_3$ & Defect phonon side band & 1.78 & --- & ---\\
\colrule
$SG_4$ & Defect trap state mediated transition & 2.01 & 2.03 & ---\\
\colrule
$DB_1$ & Direct $VB_{Y-D,\space 1} \rightarrow CB_{Y-D,\space 1}$  & 2.34 & 2.38 & 2.33\\
\colrule
$DB_2$ & Direct $VB_{\Gamma-Y,\space 1} \rightarrow CB_{\Gamma-Y,\space 1}$  & 2.59 & 2.53 & 2.57\\
\colrule
$DB_3$ & Direct $VB_{M,\space 1} \rightarrow CB_{M,\space 3}$  & 2.88 & 2.80 & 2.82\\
\colrule
$DB_4$ & Direct $VB_{M,\space 3} \rightarrow CB_{M,\space 5}$  & 3.13 & 3.02 & 3.03\\
\colrule
$IB_1$ & Indirect $VB_{M,\space 3} \rightarrow CB_{E,\space 9}$  & --- & 3.35 & 3.40\\
\colrule
$IB_2$ & Indirect $VB_{E,\space 3} \rightarrow CB_{M,\space 5}$  & --- & 3.71 & 3.73\\
\end{tabular}
\end{ruledtabular}
\caption{Identified transitions and peak energies obtained from fitting PL, $\Delta R/R$, and PR spectra. Direct interband (DB) and indirect interband (IB) transitions are matched to DFT predicted band structures of Ref.[\cite{oliveira_2d_2023}]. See  Figs. \ref{fig:supfig2} and \ref{fig:supfig3}.}
\end{table*}

\subsection{\label{bandgap} Direct/Indirect interband (DB/IB) and sub-bandgap (SG) transitions}

Taken together, the PL, $\Delta R/R$, and 1PA PR spectral measurements give a comprehensive characterization of the optically induced electronic transitions that take place in $CuScP_2S_6$ at NIR, visible, and NUV wavelengths. Two strong broad sub-$E_g$ PL emission peaks are observed at 1.32 eV and 2.01 eV, energies that correspond closely with weak $\Delta R/R$ peaks at 1.03 eV and 1.3 eV ($SG_1$ and $SG_4$). Their strong emissive and  weak reflective character paired with their non-detection in the PR spectrum along with their occupying energies below previous DFT predictions of $CuScP_2S_6$'s $E_g$ (Bulk: 2.23 eV and monolayer: 2.38 eV) \cite{oliveira_2d_2023} indicates that they likely arise from trap state mediated electronic transitions or self-trapped excitonic transitions. Self-trapped excitons have recently been reported in the structurally similar material $CuInP_2S_6$. \cite{yan_self-trapped_2026}  To obtain a close fit to the PL spectrum, two other lower intensity peaks are fitted at 1.45 eV and 1.78 eV ($SG_2$ and $SG_3$) and could be related to non-radiative recombination associated with other nearby transitions. 

Four optical transitions are identified as occurring at comparable energies in all three measurements. Among the three spectra, these transitions are measured as occurring in the following ranges- 2.34-2.38 eV, 2.53-2.59 eV, 2.80- 2.88 eV, and 3.02-3.13 eV. The fact that these transitions show optical recombination as measured via PL suggests that they are direct band to band transitions, as such these transitions are labeled $DB_1-DB_4$, respectively. The weaker emission of these peaks in the PL spectrum can be explained as occurring due to quenching effects related to the capture of free carriers by trap states whose aforementioned associated PL peaks show much larger emission intensity. The same trapping mechanisms decrease carrier diffusion length leading to a weaker associated PR as well. Two additional higher energy electronic transitions are identified in the $\Delta R/R$ and 1PA PR spectra, measured in the ranges- 3.35- 3.40 eV and 3.71-3.73 eV. The absence of PL from these transitions suggests that they may be indirect band to band transitions leading to them being assigned the labels, $IB_1$ and $IB_2$, respectively. Moreover, their enhanced PR could be attributed to the involvement of an increased contribution of hot conduction band electrons and/or hot valence band holes. In order to obtain a close fit to the $\Delta R/R$ spectrum, a higher energy peak is assumed to lie outside of the limits of the spectral range obtainable with our setup for the $\Delta R/R$ measurement.

\subsection{\label{2PAexp}Two-photon Absorption Photocurrent}

A vertical geometry photocurrent device is fabricated by making graphite- $CuScP_2S_6$ – graphite stacks via PDMS-assisted dry transfer and using maskless photolithography to deposit contacts on the top and bottom graphite layers with evaporated Ti/Au (5 nm/ 800 nm) stacks (Figure \ref{fig:fig1}c- left). AFM measurements show the $CuScP_2S_6$ layer to be $\approx$ 90 nm. The vertical device geometry is essential for measurement of BPVE photocurrents in $CuScP_2S_6$ as its structure displays an out of plane breaking of inversion symmetry due to non-symmetric Cu site occupation. \cite{siebenaller_second_2024, siebenaller_nonlinear_2025} IV measurements show the formation of asymmetric Schottky junctions at the graphite- $CuScP_2S_6$ interfaces. (Figure \ref{fig:fig1}d- right) The asymmetry likely arises due to differences in the top and bottom contact area. Similar to the lateral device, an initially slow rise in current is observed in the IV measurement for both forward and reverse sweep directions which can similarly be attributed to charging and discharging effects associated with deep level trapping states. 

\begin{figure*}
    \centering
    \includegraphics[width=5.5in]{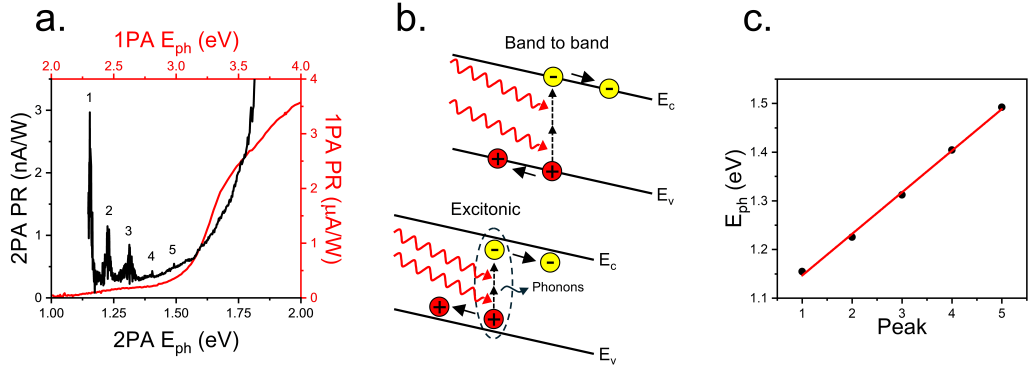}
    \caption{a.) 2PA PR spectrum obtained from $CuScP_2S_6$ lateral device under 3V bias, 1PA PR is presented as well in light red for comparison, b.) schematic illustrating band to band 2PA process "top" and excitonic 2PA processes "bottom", and c.) plot of peak energy vs transition peak number. }
    \label{fig:fig2}
\end{figure*}

The 2PA PR spectrum is collected in the lateral device following the 3 V bias collection scheme discussed in the previous section. To provide higher photon fluence, a wavelength tunable, pulsed Ti- Sapphire with a pulse length of ~140 fs (rep rate 80 MHz) was used as the excitation source. The 2PA PR spectrum is plotted next to the linear single photon absorption (1PA) PR spectrum obtained from the same device for direct comparison. (Figure \ref{fig:fig2}a) The 2PA PR spectrum shows measurable PR all the way down to $E_{ph}$=1.15 eV (1080 nm) which constitutes the upper wavelength limit of the laser and corresponds to a total transition energy of 2.3 eV. Having reached the upper wavelength limit of the laser, measurement of the entire lower band tail is not possible but the collected spectrum is consistent with the observed total cut-off of PR in the 1PA spectrum at an $E_{ph}$ of $\approx$ 2.2 eV. Furthermore, the 2PA PR spectrum shows a secondary transition edge at $\approx$ 3.0 eV that can be attributed to a higher energy transition for which the probability of 2PA is enhanced.

Distinct from the 1PA PR spectrum, three strong, narrow linewidth features, centered at 1.15 eV, 1.23 eV, and 1.31 eV, labeled peaks 1, 2, and 3, respectively, can be observed in the 2PA PR spectrum (Table 2). Two weaker peaks, labeled 4 and 5, are observed at 1.40 eV and 1.49 eV. The narrow linewidth and proximity of these peaks to the band edge is consistent with them arising from exciton transitions (Figure \ref{fig:fig2}b). Furthermore, the center $E_{ph}$'s associated with these peaks are approximately evenly spaced leading to a linear relationship between their assigned number and center $E_{ph}$ which is consistent with the peaks arising from multiple higher order phonon side band transitions (Figure \ref{fig:fig2}c). The slope of the fitted linear relationship is 85.4 $\pm$ 2.2 meV which corresponds to the energy of the involved phonon modes. Optically pumped, localized regions of high exciton dipole density are theorized to enable strong enhancements of higher order macroscopic nonlinear susceptibility including $\chi^{(3)}$ in low dimensional materials \cite{shimizu_optical_1988, hanamura_very_1988, hanamura_exciton_1989}, a phenomenon that has recently seen a plethora of reported computational predictions and experimental observations in various 2D materials. \cite{zhouTwophotonAbsorptionSubband2017,pedersen_excitonic_2021,yao_nanoscale_2022,esteve-paredes_excitons_2025,sun_exciton_2025,xu_broadband_2025} This phenomenon would help to explain the enhanced 2PA PR observed at peaks 1, 2, and 3. Moreover, this exciton could be a dark exciton that is undetectable via 1PA based techniques but becomes detectable under the more relaxed selection rules governing 2PA. \cite{bonin_two-photon_1984, birge_introduction_1984, rumi_two-photon_2010}

\begin{table}
\begin{ruledtabular}
\begin{tabular}{ccc}
Label & 2PA PR Peak Energy (eV) & Total Transition Energy (eV)\\
\colrule
\colrule
1 & 1.15 & 2.30\\
\colrule
2 & 1.23 & 2.46\\
\colrule
3 & 1.31 & 2.62\\
\colrule
4 & 1.40 & 2.80\\
\colrule
5 & 1.49 & 2.98\\
\end{tabular}
\end{ruledtabular}
\caption{Energies of 2PA PR narrow peaks labeled in Fig. \ref{fig:fig2}.}
\end{table}

The associated total transition energy of peaks 1,3, 4, and 5 is within the range of $DB_1$, $DB_3$, $DB_4$, and $DB_5$, respectively. Though it is possible that the peaks share a common source, namely the direct band to band transitions proposed in the previous section, the absence of a PL, $\Delta R/R$, or 1PA PR peak near the total radiative transition energy of 2PA PR peak 2 (2.46 eV) combined with the even energy spacing of the 2PA PR peaks as well as the side-band like decrease in intensity of progressively higher energy peaks suggests that an exciton with moderate binding energy ($E_b$) of $\approx$ 50 meV associated with the energy bands involved in $DB_1$ may be a more likely explanation. Another potentially connected feature observed in the $\Delta R/R$ is the peak edge that can be observed right at the cut off $E_{ph}$ at 1.15 eV. Since the $\Delta R/R$ spectrum was taken with a lower fluence CW light source, this peak likely corresponds to a single photon absorption transition, indicating that the 1.15 eV feature in the 2PA PR spectrum could in fact be due to 1PA photocurrent associated with some large $E_b$ exciton or deep level defect assisted transition. Assuming that $DB_1$ corresponds to the direct $E_g$ transition, this proposed exciton would display a binding energy >1.1 eV presenting a significant hurdle, barring more exotic mechanisms, to achieving the charge separation needed to produce a measureable photocurrent at the bias of 3V. Though our work cannot conclusively determine the exact origins of peaks 1-5, future computational calculations paired with gate-voltage-dependent and polarization resolved optical measurements could help to test the hypothesis that peaks 1-5 are excitonic in origin and delineate the exact nature of any underlying exciton state.\cite{cavalcante_stark_2018, katsch_excitonic_2022,hou_efficient_2025}

\subsection{\label{BPVA}Bulk Photovoltaic Effect (BPVE) Spectroscopy}

\begin{figure*}
    \centering
    \includegraphics[width=5in]{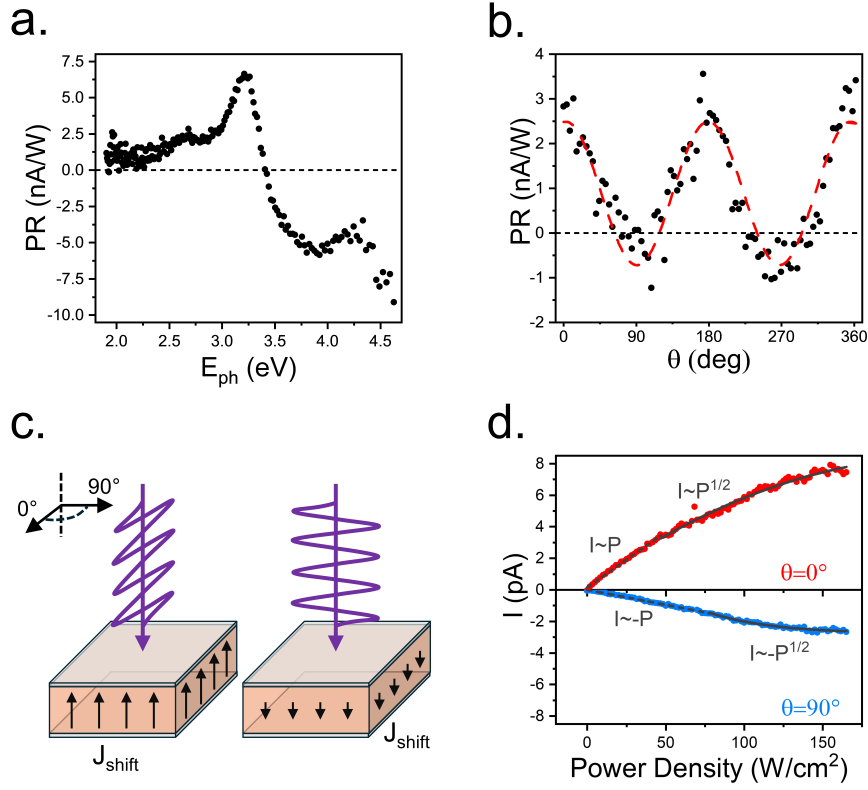}
    \caption{a.) Zero-bias BPVE PR spectrum obtained from vertical graphite-$CuScP_2S_6$-graphite device shows an anomalous sign change at 3.41 eV, b.) measurement of strong optical polarization orientation dependent PR at 405 nm fit by a sine function with period of $180$ degrees, c.) schematic diagram illustrating the polarization dependent BPVE PR, and d.) power density dependence of measured positive photocurrent excited with $\theta=0^\circ$ polarized 405 nm laser "red" and negative photocurrent excited with $\theta=90^\circ$ polarized 405 nm laser "blue" shows characteristic transition from linear to square root relationship.}
    \label{fig:fig3}
\end{figure*}

Spectrally resolved zero-bias PR measurements are obtained from the graphite-$CuScP_2S_6$- graphite vertical device to study the BPVE in $CuScP_2S_6$. (Figure \ref{fig:fig3}a) Here, positive photocurrent is defined as conventional current, i.e. positive charge, flowing from the bottom to the top contact or, equivalently, electrons flowing from the top to bottom contact. A positive PR peak centered at $\approx$ 2.65 eV is measured alongside a higher energy PR peak centered at $\approx$ 3.21 eV.  The energies of these features are consistent with transitions $DB_2$ and $IB_1$. The previously discussed PL spectrum also shows emission peaks in the vicinity of these energies. The sign of the PR flips at 3.41 eV above which PR from higher energy transitions shows a negative polarity. A negative PR peak is observed at 3.88 eV, and the magnitude of PR increases at higher $E_{ph}$ indicating a peak beyond 4.5 eV. The sign change in PR is consistent with a reversal in the sign of $\Delta\chi_{cv}$ as different $E_{ph}$ probe different regions of the Brillouin zone. Such wavelength dependent photocurrent sign changes are reported in other BPVE materials. \cite{cook_design_2017, koch_bulk_1975, young_first_2012} The magnitude of the positive PR peak centered at 3.21 eV, 6.6 nA/W, is comparable to the negative PR peak centered at 3.88 eV, -5.7 nA/W, making the observed sign change difficult to explain when solely accounting for possible photocurrent contributions originating from the Schottky contact regions.

A linearly polarized 405 nm ($E_{ph}$=3.06 eV) laser diode is paired with a $\lambda/2$ wave-plate and used to collect the polarization orientation dependent PR obtained by rotating the optical polarization axis within the ab plane by $\theta$. (Figure \ref{fig:fig3}b) The data is tightly fit by a sinusoidal function with a period of $180^\circ$, mirroring the two fold symmetry observed in previous reports of polarization orientation dependent SHG measurements in $CuScP_2S_6$.\cite{siebenaller_second_2024, siebenaller_nonlinear_2025} Both phenomena are likely linked to two fold symmetry of the Cu sub-lattice within the a-b plane. A sign change is observed, consistent with the polarization tunable $\nabla_k \varphi^{a}_{vc}$ component of the shift vector. (Figure \ref{fig:fig3}c) The increased PR at intervals of $180^\circ$ (2.5 nA/W) as compared to at intervals of $90^\circ$ (0.7 nA/W) indicates an enhanced $R$ oriented out of plane occurring for the $180^\circ$ orientation. Similar polarization dependent BPVE responses are reported in other materials. \cite{abdelsamie_crossover_2022, liang_strong_2023, zhou_symmetry-breaking-engineered_2025}

In order to further investigate the nature of the polarization induced BPVE polarity switching in $CuScP_2S_6$, optical power dependent photocurrent measurements are conducted with the 405 nm laser set to the $\theta=0^\circ$ and then the $\theta=90^\circ$ polarization states. A clear non-linear BPVE photocurrent power dependence is observed for both optical polarization states.(Figure \ref{fig:fig3}d) A transition from linear power dependence ($I\propto P$) to square-root power dependence $I\propto P^{1/2}$ is observed and fit analytically. Such a transition occurs in other BPVE materials.\cite{zhang_switchable_2022,zhou_giant_2024, liang_strong_2023, qiao_boosting_2025, zhou_cd_2026} In these cases the shift to $I\propto P^{1/2}$ power dependence is thought to occur due to quadratic recombination processes becoming the limiting factor in shift photocurrent generation. Interestingly, the photocurrent generated under the $\theta=0^\circ$ polarization state shows a transition to $I\propto P^{1/2}$ power dependence at a lower power density. This fact can be attributed to the enhanced BPVE response associated with the $\theta=0^\circ$ polarization orientation causing the shift current to become recombination limited at lower optical power. Together, the spectral, polarization, and power dependence measurements of photocurrent in our vertical $CuScP_2S_6$ device offer solid evidence for a polarization sensitive BPVE, though further theoretical studies are required to clarify the precise origin of this observed phenomenon. 

\section{\label{sec:concl} Conclusions}

The linear optical and optoelectronic properties of $CuScP_2S_6$ show it to be a wide-band gap semiconductor in agreement with the computed electronic structure of Oliveira \textit{et al.}~\cite{oliveira_2d_2023}.  Furthermore, this material displays non-linear optoelectronic phenomena, including 2PA and BPVE photocurrent. Spectrally-resolved PR shows strong 2PA photocurrent and a series of sharp excitonic 2PA PR peaks separated by 85 meV indicative of phonon-mediated transitions. BPVE driven photocurrent measurements show a wavelength dependent sign change and strong optical polarization orientation dependence consistent with changes in the shift vector. Taken together, this demonstrates $CuScP_2S_6$ to be a candidate platform for ultrafast 2PA-enabled near-infrared photodetection and polarization-sensitive photodetection in the UV-VIS. The devices studied here all utilized sub-optimal back-to-back Schottky contacts. Future improvements in defect control in $CuScP_2S_6$ and Ohmic contact development should therefore greatly bolster its non-linear photoresponsivity and elevate its potential for applications in these areas.

\section{Acknowledgments}
This research was supported by a Catalyst Grant from The Ohio State University President’s Research Excellence (PRE) program. Work performed at AFRL was financially supported by the U.S. Air Force Office of Scientific Research LRIR Grant No. RX26COR010. The authors acknowledge the use of facilities and instrumentation at the Nanotech West Laboratory, a core user facility managed by the Institute for Materials and Manufacturing Research (IMR) at The Ohio State University.

\section{Data Availability}
The data that support the findings of this article are not publicly available. The data are available from the authors upon reasonable request.

\appendix*

\setcounter{figure}{0} 
\renewcommand{\thefigure}{A\arabic{figure}} 

\section{Supplementary Data, Analysis, and Calibrations}
\subsection{\label{sec:raman} Raman Spectroscopy}

\begin{figure}[h]
    \centering
    \includegraphics[width=2.5in]{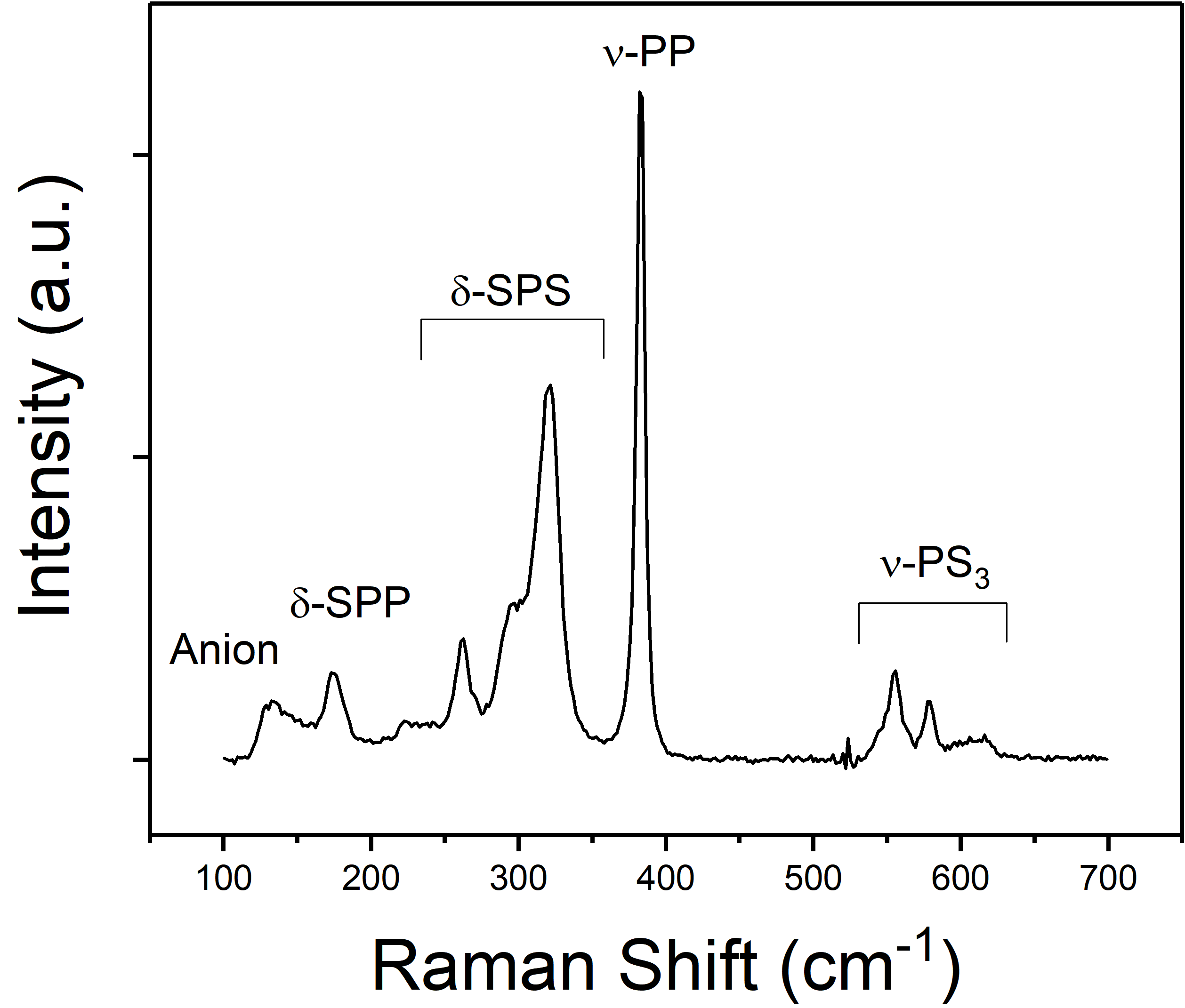}
    \caption{Raman spectrum of $CuScP_2S_6$ measured using a 514 nm excitation laser. The various Raman active modes are labeled in the figure.}
    \label{fig:supfig1}
\end{figure}

Figure \ref{fig:supfig1} presents the Raman spectrum collected in order to characterize the $CuScP_2S_6$ sample. Here, the Si substrate Raman peak at $520\ cm^{-1}$ was fitted using a Voigt peak profile in Python and subtracted during post-processing in RamanSPy \cite{Dimitar2024} to present the spectrum due to the sample itself. The various Raman active modes correspond to the anion translations and librations, as well as the $\delta$-SPS and the $\nu$-$P_2S_6$ stretching modes, as labeled in the figure. These match well with the characteristic Raman modes previously noted for $CuScP_2S_6$ in literature \cite{ghosh_light-induced_2026, oliveira_2d_2023}, thus characterizing the sample, and firmly establishing its identity.

\subsection{\label{sec:exp-vs-dft} Correlation of Experimentally Measured Peaks with Previously Published DFT Calculated Band Structure and Density of States}

Fitted peak energies found in Table 1 were compared with Oliveira \textit{et al.}’s DFT calculated band structure and density of states for $CuScP_2S_6$.\cite{oliveira_2d_2023} Transitions that displayed PL emission “$DE_1$- $DE_4$" are characterized as likely to be direct in nature and corresponding direct band to band transitions are identified on the band diagram (Fig. \ref{fig:supfig2}). Similarly, transitions for which no PL emission was measured are assumed to be indirect and corresponding indirect band to band transitions are identified on the band diagram. (Fig. \ref{fig:supfig3}) Transition assignments are intended to be preliminary. Future calculations that probe transitions probability, e.g. DFT derived transition dipole moments and joint density of states (JDOS) calculations, of these and other potential transitions could provide more definitive evidence to assign specific transitions to the experimentally observed peak energies.

\begin{figure*}[h]
    \centering
    \includegraphics[width=6.75in]{Supplementary_Figure_2.png}
    \caption{Direct band to band transitions identified in DFT calculated band structure and density of states from Oliveira \textit{et al.} that are consistent with peak energies fitted in experimentally measured PL, $\Delta R/R$, and PR spectra.\cite{oliveira_2d_2023}}
    \label{fig:supfig2}
\end{figure*}

\begin{figure*}[h]
    \centering
    \includegraphics[width=6.75in]{Supplementary_Figure_3.png}
    \caption{Indirect band to band transitions identified in DFT calculated band structure and density of states from Oliveira \textit{et al.} that are consistent with peak energies fitted in experimentally measured PL, $\Delta R/R$, and PR spectra.\cite{oliveira_2d_2023}}
    \label{fig:supfig3}
\end{figure*}

\subsection{\label{sec:calibration} Characterization and Calibration of Light Sources Used for PR Measurements}
\begin{figure*}[h]
    \centering
    \includegraphics[width=6.75in]{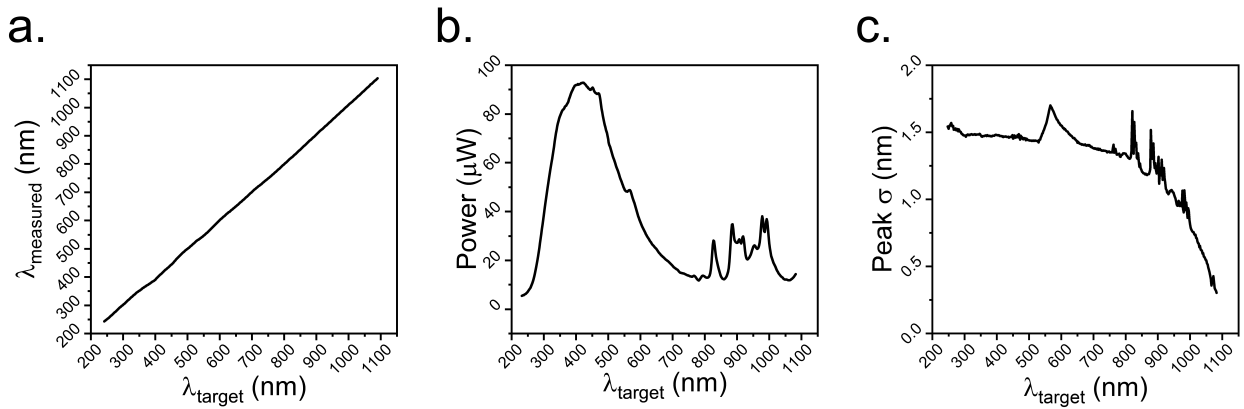}
    \caption{a. Plot relating targeted monochromator wavelength and actual fitted center wavelength of monochromated light spectrum at wavelengths used for photocurrent measurement, b. standard deviation associated with monochromated Gaussian spectral peaks, and c. power spectrum of wavelength calibrated light source.}
    \label{fig:supfig4}
\end{figure*}

\begin{figure*}[h]
    \centering
    \includegraphics[width=4.5in]{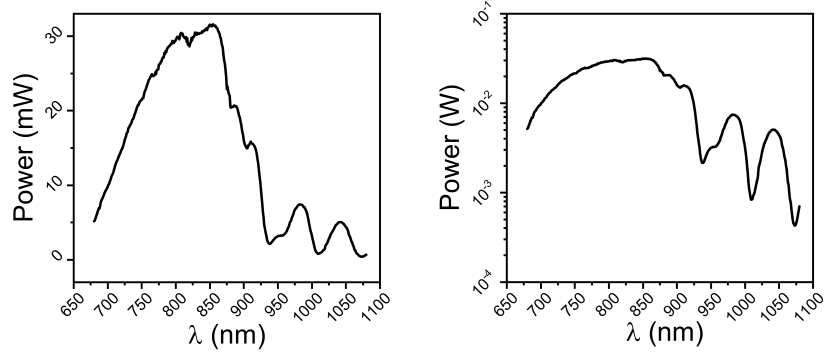}
    \caption{Measured power spectrum of wavelength tunable Ti-sapphire laser. A log-scaled plot of the spectrum is included to better relate the power in the lower power region of the spectrum.}
    \label{fig:supfig5}
\end{figure*}

A spectrometer is used to measure the optical spectrum of the monochromated light source for each target wavelength that the photocurrent measurement was conducted at (248 nm- 1080 nm with 2 nm steps). Each spectrum consists of a Gaussian profile centered at a wavelength near the target wavelength. A Gaussian fit routine is applied to every spectrum and used to extract the center wavelength of the associated target wavelength which is then used to precisely calibrate the wavelength “$E_ph$" axis of the PR measurements (Fig. \ref{fig:supfig4}a) The standard deviation ($\sigma$) associated with each calibrated wavelength is also extracted and found to be <1.75 nm at all wavelengths. (Fig. \ref{fig:supfig4}b) Lastly, an optical power meter is used to measure the power at each calibrated wavelength used for the PR measurements. (Fig. \ref{fig:supfig4}c) The optical power meter is placed at the location of the sample after all optics to ensure that power is measured under experimental conditions. Similarly, the optical power of the wavelength tunable Ti-Sapphire laser is measured at the wavelengths used for the 2PA PR measurement (680 nm- 1080 nm with 2 nm steps), see Fig. \ref{fig:supfig5}.

To characterize the polarization purity of the polarized laser used for the polarization dependent photocurrent measurements, a Glan-Thompson calcite polarizer is placed in front of the laser diode, clean up plate polarizer, and $\frac{\lambda}{2}$ wave plate. A power meter is placed after a 40x reflective objective to measure the optical power. The wave plate is rotated using a piezo actuated automatic rotation stage to obtain the lowest possible power at the power meter, 6.48E-8 W, constituting the position at which the polarized light has been rotated to be exactly orthogonal to the calcite polarizer analyzer. Then a control program based on Malus’s law is used to rotate the wave plate at specific angles calculated to give a constant increase in power with each, i.e. linearizing the power to the step number. A linear fit routine is employed resulting in a fit with a residual sum of squares of 1.4E-11, indicating that the extinction position of the polarizer has been determined with ultrahigh precision.  The maximum power, 5.83E-4 W, is obtained for the position corresponding to the polarization state of the light aligning parallel to the polarizer analyzer. Thus a polarization extinction ratio of 9000:1 is measured.  

\begin{figure*}[h]
    \centering
    \includegraphics[width=4.5in]{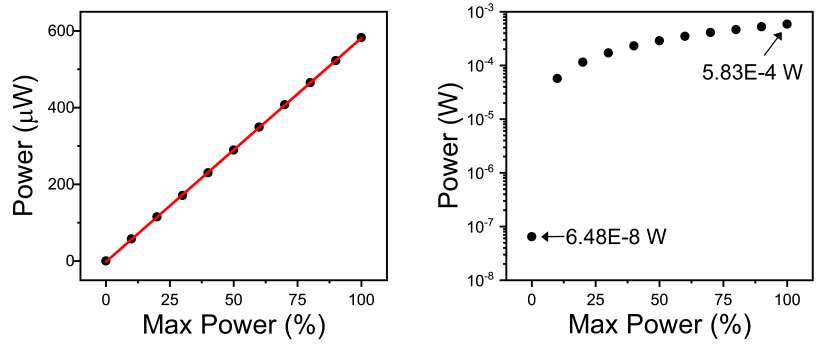}
    \caption{Measurement of laser power as function of max power percentage. A Malus’ Law based program is used to calculate the precise polarizer angles needed for each power percentage. Log-scaled data is included to show power at extinction angle.}
    \label{fig:supfig6}
\end{figure*}

\bibliography{aipsamp}

\end{document}